\documentclass[usegraphix,usenatbib,fleqn,useAMS]{mn2e}

\usepackage{hyperref}
\usepackage{amsmath}
\usepackage{graphicx}
\usepackage{subfigure}
\usepackage{epsfig}
\usepackage[varg]{txfonts}

\renewcommand\le\oldleq
\renewcommand\ge\oldgeq
\renewcommand\pi\upi

\def\dm{{\rm d}}
\def\im{{\rm i}}
\def\rme{{\rm e}}

\begin{document}


\title[The Doubloon Models]
    {The Doubloon Models}

\author[Evans et al.]
    {N. W. Evans$^1$\thanks{E-mail: nwe,adb,aamw3@ast.cam.ac.uk, jinan@nao.cas.cn},
     J. An$^2$, A. Bowden$^1$, A. A. Williams$^1$
\medskip
   \\$^1$Institute of Astronomy, University of Cambridge, Madingley Road,
         Cambridge, CB3 0HA, UK
    \\$^2$National Astronomical Observatories,
Chinese Academy of Sciences, A20 Datun Road, Chaoyang District,
Beijing 100012, PR~China
  }
\maketitle

\begin{abstract}
A family of spherical halo models with flat circular velocity curves
is presented. This includes models in which the rotation curve has a
finite central value but declines outwards (like the Jaffe model). It
includes models in which the rotation curve is rising in the inner
parts, but flattens asymptotically (like the Binney model). The family
encompasses models with both finite and singular (cuspy) density
profiles.  The self-consistent distribution function depending on
binding energy $E$ and angular momentum $L$ is derived and the
kinematical properties of the models discussed. These really describe
the properties of the total matter (both luminous and dark).

For comparison with observations, it is better to consider tracer
populations of stars. These can be used to represent elliptical
galaxies or the spheroidal components of spiral galaxies. Accordingly,
we study the properties of tracers with power-law or Einasto profiles
moving in the doubloon potential. Under the assumption of spherical
alignment, we provide a simple way to solve the Jeans equations for
the velocity dispersions. This choice of alignment is supported by
observations on the stellar halo of the Milky Way.  Power-law tracers
have prolate spheroidal velocity ellipsoids everywhere. However, this
is not the case for Einasto tracers, for which the velocity ellipsoids
change from prolate to oblate spheroidal near the pole.

Asymptotic forms of the velocity distributions close to the escape
speed are also derived, with an eye to application to the high
velocity stars in the Milky Way. Power-law tracers have power-law or
Maxwellian velocity distributions tails, whereas Einasto tracers have
super-exponential cut-offs.
\end{abstract}

\begin{keywords}
galaxies: haloes -- galaxies: kinematics and dynamics -- stellar
dynamics -- dark matter
\end{keywords}

\section{INTRODUCTION}

The study of spherical models is useful both for representing galaxies
and dark haloes. Even though flattening is often important, spherical
models have provided useful insights into the behaviour of stellar
systems. They can serve as starting points for more flattened models,
either as initial conditions for N-Body experiments or as the lowest
order terms in basis function expansions \citep{He92}.

An important family of spherical models discovered over the last
twenty-five years is the double-power law or $\gamma$ models
\citep{De93, Tr94}. These have a density profile that is cusped like
$r^{-\gamma}$ at small radii, yet falls like $r^{-4}$ at large radii.
They have found ready applications in modern astronomy. They include
two particularly simple and appealing models found earlier by
\citet{Ja83} and \citet{He90}, which differ in the strength of the
central density cusp, $\rho \sim r^{-2}$ and $r^{-1}$
respectively. All the double-power law models generate simple
gravitational potentials or force-laws.  They also have analytically
tractable distribution functions (DFs). This is technically
challenging, as an integral equation has to be solved \citep[see
  e.g.,][]{Ed16,BT}. Nonetheless, DFs are useful as they encode all
the properties of the model, enabling initial conditions to be set for
N-body realisations of distributions of observables to be computed.

Here, we provide another very simple family of spherical double-power
law halo models -- {\it the doubloon models}. They are motivated by
the flatness of galaxy rotation curves, and so the models have a
density that falls like $r^{-2}$, either in the inner parts or the
outer parts. This generates a regime in which the rotation curve is
roughly flattish.  Explicitly, we consider the potential-density pair
\begin{equation}
  \psi = -{V^2 \over p} \log \biggl( {r^p+a^p \over a^p} \biggr),
  \quad
  \rho = {V^2 \over 4\pi G r^{2-p}}
         {(1+p)\, a^p+r^p \over (a^p+r^p)^2}.
  \label{eq:doubloon}
\end{equation}
Then, the rotation curve of the model is
\begin{equation}
  v_{\rm c} (r)
  = V \biggl( 1 + {a^p \over r^p} \biggr)^{-1/2}.
\end{equation}
Here, $V$ is the amplitude of the flat rotation curve, whilst $a$ is a
scalelength.  The density is positive everywhere provided $p\ge -1$,
whereas it is monotonically decreasing (and hence astrophysically
realistic) provided $p\le 2$.  When $p=0$, the model degenerates into
the singular isothermal sphere (i.e.\ $\rho\propto r^{-2}$ but the
strict limit is scaled to $v_{\rm c}=V/\sqrt2$), known for over a
century thanks to the labours of J. H. Lane and R. Emden \citep[see
  also][]{Ch10}.

The properties of the family divide neatly into two, as shown by their
central and asymptotic behaviour.  If $p>0$, the density falls off
like $r^{-2}$ as $r\to\infty$ and the rotation curve is flat
asymptotically
\begin{equation}
  v_{\rm c} (r) \simeq V \times
    \begin{cases} (r/a)^{p/2} & (r \ll a)      \\
                   1          & (r \gg a) \end{cases},
\end{equation}
which we shall subsequently refer to the {\it outer branch}.
The potential of the outer branch decreases from $\psi(0)=0$ to
$\psi(\infty)=-\infty$ as $r$ increases, and the models possess
the central density cusp like $1/r^{2-p}$ (except for $p=2$).
Some of the usual suspects are represented in the family. When $p=2$,
the model is the spherical limit of Binney's logarithmic potential
\citep{Bi81, Ev93, BT}. When $p=1$, the model has a $1/r$ cusp and has
recently been discussed by \citet{Ev14}.

The rotation curve for models with $p<0$ tends to a finite value at
the centre
\begin{equation}
  v_{\rm c} (r) \simeq V \times
    \begin{cases} 1 & (r \ll a) \\
      (a/r)^{|p|/2} & (r \gg a) \end{cases},
\end{equation}
but falls off like $r^{-|p|/2}$ as $r\to\infty$. Henceforth these
models will be referred to as the {\it inner branch}.  The density
always has an isothermal cusp ($\sim r^{-2}$) at the centre and decays
like $r^{-(2+|p|)}$ as $r\to\infty$ (with the exception of the case
$p=-1$), whereas the potential runs from $\psi(0)=\infty$ to
$\psi(\infty)=0$. The model with $p=-1$ and was introduced by
\citet{Ja83} as a representation of elliptical galaxies and has finite
mass ($M=aV^2/G$). The remaining members of the family have rotation
curves which fall off less steeply than the Jaffe model.

Both branches of the doubloon family are useful. For example, models
on the inner branch are helpful in studies of the outer parts of
galaxies, when the flat rotation curve gives out and the density of
the dark matter begins to fade. Evidence, for example, from the
Sagittarius stream suggests that the rotation curve of the Milky Way
is flat out to $\sim 50~\mbox{kpc}$, and then begins to fall \citep{Gi14}.
Models on the outer branch are useful for studying the inner parts of
galaxies, in the regime of the flat rotation curve. Their central
cusps make them desirable models of dark matter haloes, in accord with
predictions of dissipationless theories of galaxy
formation \citep{Mo10}.

The paper is arranged as follows. Section 2 studies the
self-consistent model, and gives families for the distribution
functions for the total (dark and luminous) matter.  These are
self-consistent models and so are useful both for setting up the
initial conditions for N-body experiments and for studying the
kinematic properties of the dark and luminous matter. The remainder of
the paper studies tracer populations moving in the doubloon potential.
The tracers might represent stellar populations in elliptical galaxies
or the spheroidal components of spiral galaxies. In particular,
Section 3 looks at the Jeans solutions of tracer populations (with
power-law or Einasto profiles), whilst Section 4 studies the
distributions of high velocity stars. Both applications are motivated
by the data on halo stars in the Milky Way, which has increased in
quality and quantity in recent years thanks to surveys like SDSS and
RAVE~\citep[see e.g.,][]{Sm07, Sm09a, Sm09b, Bo10, Pi14}.

\begin{figure}
  \centering
  \includegraphics[width=3in]{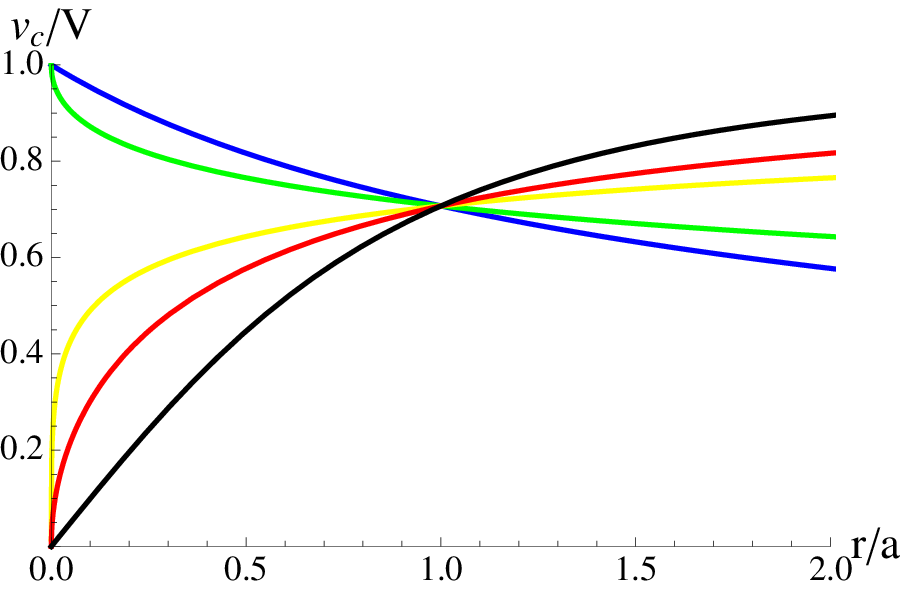}
  \includegraphics[width=3in]{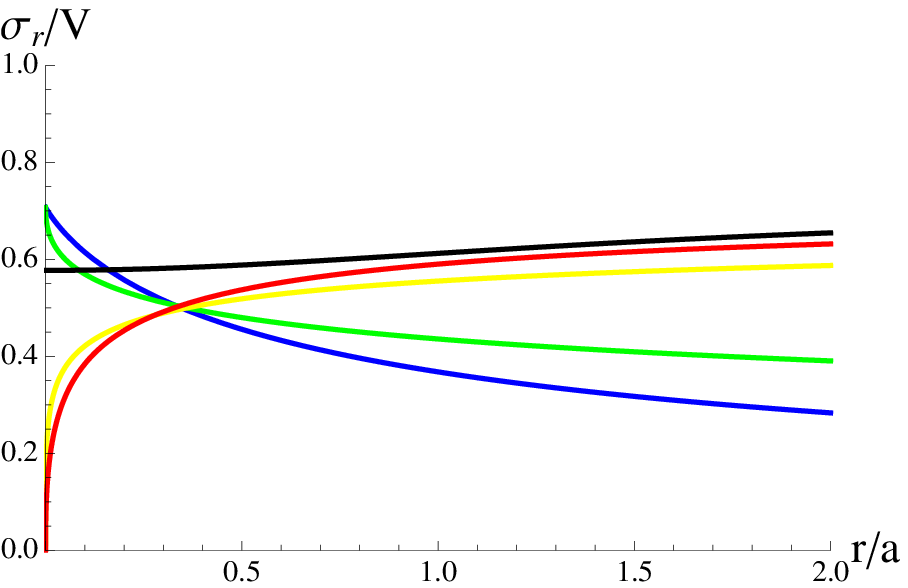}
  \caption{Rotation curves $v_{\rm c}$ (upper) and isotropic velocity
    dispersion $\sigma_r$ profiles (lower) for the models $p=2$
    (black), $p=1$ (red), $p =1/2$ (yellow), $p=-1/2$ (green) and
    $p=-1$ (blue).}
  \label{fig:velprops}
\end{figure}

\section{The Self Consistent Model}

The phase space distribution function (DF) may depend on the integrals
of motion only, as first realised by \citet{Je19}. For a spherical
potential, the integrals may be taken as the binding energy per unit
mass $E$ and the modulus of the angular momentum $L$, namely
\begin{equation}
  E = -\tfrac12({v_r^2 + v_\theta^2 + v_\phi^2}) + \psi,
  \quad L^2 = r^2(v_\theta^2 + v_\phi^2).
\end{equation}
Isotropic DFs depend only on $E$, whereas anisotropic DFs depend on
$L$ as well.  Here we look for constant anisotropy DFs; that is, the
anisotropy parameter $\beta \equiv 1 - \langle v_\theta^2
\rangle/\langle v_r^2 \rangle$ takes a constant value. The parameter
$\beta$ may take values in the range $-\infty \le \beta \le 1$. When
$\beta \rightarrow 1$, the model is built from radial orbits, whilst
when $\beta \rightarrow -\infty$, it is made from circular orbits.
Constant anisotropy DFs are widely-used because of their simplicity.
However, it is important to acknowledge that there is no underlying
physical justification. Simulations of the growth of galaxies suggest
that DFs are typically isotropic in the center ($\beta \approx 0$),
but become more radially anisotropic ($\beta \approx 1/2$) in the
outer parts \citep{Ha06}. It is nonetheless reasonable to expect that
constant anisotropy DFs provide good approximations in certain
regimes, such as the central parts or the outer periphery.

The velocity dispersions of the constant anisotropy models are
found by integrating the Jeans equation, namely
\begin{equation}
\langle v_r^2\rangle=\frac1{r^{2\beta}\rho(r)}
\int_\infty^r\!\dm\tilde r\,\tilde r^{2\beta}\rho(\tilde r)\,
\frac{\dm\psi}{\dm r}\biggr\rvert_{r=\tilde r},
\end{equation}
and $\langle v_\theta^2\rangle=(1-\beta)\langle v_r^2\rangle$.
For the self-consistent doubloon models in equation (\ref{eq:doubloon})
on the outer branch, assuming $0<p\le2(1-\beta)$,
\begin{multline}
\langle v_r^2\rangle = { V^2 r^p \over (1+p)\,a^p+r^p}
\\\times
\left[{ r^{2(1-\beta-p)} (a^p+r^p)^2 \over a^{2(1-\beta)} }
(1+\lambda){\rm B}_y\bigl(1+\lambda,1-\lambda\bigr)
-{1\over 2}\right],
\end{multline}
where $\lambda\equiv2p^{-1}(1-\beta)-1\ge0$ and $y\equiv
a^p/(a^p+r^p)$, whilst ${\rm B}_x(a,b)$ is the incomplete beta
function (\citealt[\S~6.6]{AS}; \citealt[\S~8.17]{DLMF}).  For all
members, the velocity dispersion tends asymptotically to $\langle
v_r^2\rangle^{1/2}\to V/\!\sqrt{2(1-\beta)}$ as $r\to\infty$.  As
$r\to0$, it behaves like $\langle v_r^2\rangle\sim
r^{\min(p,2-2\beta-p)}$, with the exception of the case $\beta=1-p$
when $\langle v_r^2\rangle\sim r^p\log r$.  The radial velocity
dispersion velocity tends to zero at the center, except for when
$\beta=1-p/2$.

For the inner branch models ($-1\le p<0$), we find
\begin{equation}
\langle v_r^2\rangle={V^2\over 1+(1-|p|)\,x}
\left[\left({r\over a}\right)^{2(1-\beta)}
(1+x)^2\xi{\rm B}_{1\over1+x}\bigl(2+\xi,-\xi\bigr)
-{1\over 2}\right],
\end{equation}
where $\xi\equiv2|p|^{-1}(1-\beta)=-1-\lambda\ge0$ and
$x\equiv(a/r)^p=(r/a)^{|p|}$. This is finite at the centre for
$\beta<1$, as $\lim_{r\to0}\langle
v_r^2\rangle^{1/2}=V/\!\sqrt{2(1-\beta)}$, whilst it decays like
$\langle v_r^2\rangle\simeq(V^2/2)(1-\beta+|p|)^{-1}(a/r)^{|p|}$ as
$r\to\infty$. Plots of the isotropic ($\beta=0$) velocity dispersion
are shown in Fig~\ref{fig:velprops} as an illustration of some of
these properties.

The key to the simplicity of the doubloon models is that $\psi(r)$ can
be easily inverted, namely
\begin{equation}
  r (\psi) = a\left[\exp \biggl(-{ p\psi \over V^2}\biggr) - 1\right]^{1/p}.
\label{eq:req}
\end{equation}
This means that $\rho(\psi)$ can also be easily constructed
  \begin{equation}
  \rho(\psi) = {V^2\over 4 \pi G a^2}
    { \rme^{2\psi/V^2} + p\, \rme^{(p+2) \psi/V^2}
    \over \bigl( 1 - \rme^{p\psi/V^2} \bigr)^{2/p-1} }.
\label{eq:dens}
  \end{equation}
From this, we can use Eddington's (1916) formula to derive the
isotropic DF via Abel transforms \citep[see][]{BT}.
For anisotropic DFs, we similarly construct the augmented density
\begin{equation}
  g(\psi) \equiv (2r^2)^\beta \rho(\psi)
          = { 2^{\beta-2} V^2 \over \pi G a^{2(1-\beta)} }
    { \rme^{p\psi/V^2} + p\, \rme^{2p \psi/V^2}
    \over \bigl( \rme^{-p\psi/V^2} -1 \bigr)^{2(1-\beta)/p-1} }.
    \label{eq:doubloonad}
\end{equation}
%
Then the constant anisotropy DF has the form of:
\begin{equation}\label{eq:cdf}
  F(E,L)=\frac{f_E(E)\,H_\beta(L^2)}{(2\pi)^{3/2}},\quad
  H_\beta(L^2)\equiv\begin{cases}
  \dfrac{L^{-2\beta}}{\Gamma(1-\beta)}&(\beta<1)\\
  \delta(L^2)&(\beta=1)\end{cases},
\end{equation}
where $\delta(x)$ is the Dirac delta function,
whilst $g(\psi)$ at a fixed $\beta$ is given as an integral transformation
of $f_E(E)$. The explicit relationship between $g(\psi)$ and $f_E(E)$ is found
in the literature; e.g., \citet[eq.~2]{We99} or \citet[eqs.~2 \& 3]{Ev06}.
We outline a general algorithm to find $f_E(E)$ given $g(\psi)$ in 
Appendix \ref{app},  which is an elaboration of Eddington's inversion
for isotropic DFs in terms of Abel transforms.

We note that the outer branch gives simpler DFs than the inner branch.
For the outer branch, the DF is always a sum over isothermal or
Maxwell--Boltzmann distributions. Usually, the sum is infinite, but,
for some special cases, the sum is finite. For the inner branch, the
DF is a sum over incomplete gamma functions
(cf.\ \citealt[vol.~2, chap.~IX]{Er53}; \citealt[\S~6.5]{AS};
\citealt[chap.~8]{DLMF}).
Under some circumstances, the sum is finite and over Dawson's integrals
(which are equivalent to error functions).  Although
we give the general solutions, we point out some of the simple cases
along the way. 

\subsection{Outer Branch}

A physical model must have an everywhere positive DF.
Since $\rho\sim r^{2-p}$ as $r\to0$ for the self-consistent model with $p>0$,
the cusp slope--central anisotropy theorem of \citet{An06} indicates
that the constant anisotropy DF is physical only if $2\beta\le2-p$.
The constant anisotropy model for $p>0$ has the DF expressible as the sum
over the exponentials of $p\bar{\mathcal E}$
where $\bar{\mathcal E}\equiv E/V^2$, namely
\begin{multline}\label{eq:cdfo}
  F=\frac{(1-\beta)^{3/2-\beta}}{4\pi^{5/2}\Gamma(1-\beta)}
  \frac{a^{2\beta-2}V^{2\beta-1}}{GL^{2\beta}}
  \rme^{(1+\lambda)p\bar{\mathcal E}}
    \\\times
    \Biggl[1+\sum_{j=1}^\infty
    \biggl(\frac{(\lambda-1)_j}{j!}+\frac{(1+p)(\lambda)_{j-1}}{(j-1)!}\biggr)
    \biggl(1+\frac{j}{1+\lambda}\biggr)^{\frac32-\beta}
    \rme^{jp\bar{\mathcal E}}\Biggr],
\end{multline}
where $(\lambda)_j = \prod_{i=0}^{j-1}(\lambda+i)$
is the Pochhammer symbol and $\Gamma(x)$ is the gamma function.
Also note $\lambda\equiv2(1-\beta)/p-1\ge0$.
For $p=1$ (the halo model with the $1/r$ density cusp), this
reduces to the expression given by \citet{Ev14}.  Experimentation
shows that the sum over the exponentials converges rapidly, so
this is a practical way to compute the DF.

There is a particularly simple case that is worthy of note. If
$\lambda=0$ (i.e.\ $\beta=1-p/2$), then
the DF reduces to just a sum over two isothermals, multiplied by a
power of the angular momentum
\begin{equation}
  F=\frac{V^{1-p}L^{p-2}}{4\pi^{5/2}\Gamma(p/2)\,Ga^p}\,
  \biggl[\left({p\over 2}\right)^{\frac{1+p}2}\rme^{p\bar{\mathcal E}}
    +p^{\frac{3+p}2}\rme^{2p\bar{\mathcal E}}\biggr].
\label{eq:simpleDF}
\end{equation}
These are amongst the simplest DFs known, and some particular cases of
this family have already appeared in the literature. So, for Binney's
logarithmic model ($p=2$),
\begin{equation}
  F=\frac1{\pi^{5/2}Ga^2V}
  \Biggl[\frac14\exp\biggl(\frac{2E}{V^2}\biggr)
  +\!\sqrt2\exp\biggl(\frac{4E}{V^2}\biggr)\Biggr],
\label{eq:isodf_bl}
\end{equation}
which reduces to the isotropic DF comprised of two isothermals given
in \citet{Ev93}. For the halo model with the $1/r$ density cusp
($p=1$), it reduces to the simple anisotropic DF found in
\citet{Ev14},
\begin{equation}
  F=\frac1{\pi^3GaL}
  \Biggl[\frac18\exp\biggl(\frac{E}{V^2}\biggr)
  +\frac14\exp\biggl(\frac{2E}{V^2}\biggr)\Biggr].
\end{equation}
For all $p$, the velocity dispersions corresponding to the DF of
equation (\ref{eq:simpleDF}) are analytic and given by:
\begin{equation}
  \langle v_r^2 \rangle = {V^2\over 2p}{(2+p)\,a^p+2r^p\over
    (1+p)\,a^p+r^p },\quad
  \langle v_\theta^2 \rangle = 
  {V^2\over 4}{(2+p)\,a^p+2r^p\over (1+p)\,a^p+r^p }.
\end{equation}
This provides a simple solution to the Jeans equations for the
velocity dispersions.

We remark that (i) the same simplicity occurs for the hypervirial
models\footnote{Hypervirial models satisfy the virial theorem
  locally. The most celebrated example is the \citet{Pl11} model, for
  which the property of hyper-viriality was established by
  \citet{Ed16}.  \citet{Ev05} found a family of models with this
  property -- See also \citet{An06b}. The property of hyperviriality
  has been studied theoretically by \citet{Ig06} and \citet{So08}}
when the cusp slope $p$ and anisotropy $\beta$ are related
$\beta=1-p/2$ \citep{Ev05} and that (ii) this combination of cusp
slope and anisotropy is at the extreme permitted by the cusp
slope-anisotropy theorem.

If $\lambda=1$ (i.e.\ $\beta=1-p$) on the other hand,
\begin{equation}
  F=\frac{p^{1/2+p}V^{1-2p}L^{2(p-1)}}
  {4\pi^{5/2}\Gamma(p)\,Ga^{2p}}
 \biggl(\rme^{2p\bar{\mathcal E}}
    +(1+p)\sum_{j=3}^\infty\left({j \over 2}\right)^{\frac12+p}
    \rme^{jp\bar{\mathcal E}}\biggr).
\end{equation}
with the velocity dispersions expressible analytically to be
\begin{multline}
 \langle v_r^2\rangle = {V^2r^p\over (1+p)\,a^p+r^p}
 \\\times
 \left\lbrace {1+p\over p} \Biggl[
  \biggl(1+{r^p\over a^p}\biggr)^2\log\biggl(1+{a^p\over r^p}\biggr)
  -{r^p\over a^p} \Biggr] - {2+3p\over 2p} \right\rbrace.
\end{multline}

Finally, we give the isotropic ($\beta=0$ and so $\lambda=2/p-1$) or ergodic DF that depends only on
energy, which is, for $0<p\le2$
\begin{multline}
 F = { 1 \over 4 \pi^{5/2} Ga^2V}\exp\biggl(\frac{2E}{V^2}\biggr)\,
 \\\times
 \Biggl[ 1 + \sum_{j=1}^{\infty}
  \biggl(\frac{({2\over p}-2)_j}{j!}
  +\frac{(1+p)({2\over p}-1)_{j-1}}{(j-1)!}\biggr)
  \biggl(1 + {jp \over 2}\biggr)^{\frac32}
 \rme^{jp\bar{\mathcal E}}\Biggr].
\label{eq:isodf}
\end{multline}
For $p=2$, this simplifies to equation (\ref{eq:isodf_bl}),
whilst if $p=1$, this reduces to
\begin{equation}
 F = { 1 \over 4 \pi^{5/2} Ga^2V}\,
 \Biggl[ \exp\biggl(\frac{2E}{V^2}\biggr) +   (1+p)\sum_{j=3}^{\infty}
  \biggl({j \over 2}\biggr)^{\frac32}\exp\biggl(\frac{jE}{V^2}\biggr)\Biggr].
\end{equation}
The isotropic velocity dispersion resulting from the ergodic DF is
\begin{equation}
 \langle v_r^2\rangle =
 { V^2 r^p \over (1+p)\,a^p + r^p}
\left[{r^{2(1-p)}(a^p+r^p)^2 \over a^2}
{2\over p}{\rm B}_y\biggl(\frac2p,2-\frac2p\biggr)
 - {1\over2} \right]
\end{equation}
where $y\equiv a^p/(a^p+r^p)$, with the particular cases of
\begin{equation}\begin{split}
 \langle v_r^2\rangle &= {V^2\over 2}
  {2a^2+r^2 \over 3a^2+r^2} & (p=2),
\\
 \langle v_r^2\rangle &= {V^2r\over 2a+r}
  \biggl[ {2(a+r)^2\over a^2} \log \Bigl(1+{a\over r}\Bigr)
   -{5a+4r\over 2a} \biggr] &(p=1).
\end{split}\end{equation}
As $r \to 0$, the velocity dispersion tends to zero (unless $p=2$,
for which $\langle v_r^2\rangle\to V^2/3$),
whilst it tends to $\langle v_r^2\rangle\to V^2/2$ as $r\to\infty$.

\subsection{Inner Branch}

The inner branches possesses more complicated DFs than the outer
branch. This may be guessed from the properties of the \citet{Ja83}
model, whose isotropic DF, first found by Jaffe and subsequently
reported in \cite{BT}, is already a sum of special functions
(viz.\ Dawson's integral).

Once we expand $g(\psi)$ in a power-series in $\rme^{-|p|\psi/V^2}\le
1$, the constant anisotropy DF in general is expressible as a sum over
the incomplete gamma functions. The sum terminates after a finite
number of terms if $2|p|^{-1}(1-\beta)$ is a non-negative integer. If
$\beta$ is a half-integer, the actual operation reduces to ordinary
derivatives (rather than Abel transforms or fractional derivatives)
and so the result is eventually expressible by means of a rational
function of exponentials. If $\beta$ is an integer, the resulting
incomplete gamma functions are reducible to the error function (or
equivalently Dawson's integral).  If $|p|=2(1-\beta)/n\le1$ where $n$
is a positive integer, the final expression is resolved into a finite
sum over such functions, as is the case for the isotropic Jaffe model.

\subsubsection{Models with radially biased orbit distributions}

The simplest constant anisotropy DFs for the inner branch models are
obtained for $\beta=1/2$. These models are of widespread physical
applicability, as numerical simulations suggest that these are
characteristic of dark matter haloes, at least in the outer parts
\citep[e.g.,][]{Ha06}.  Using the notation $\bar{\mathcal E}\equiv
E/V^2$, we find:
\begin{equation}
 F=\frac{\bigl(\rme^{|p|\bar{\mathcal E}}-1\bigr)^{1/|p|}}{(2\pi)^3GaL}
 \left(1+|p|^2\rme^{-|p|\bar{\mathcal E}}
 -2|p|^2\rme^{-2|p|\bar{\mathcal E}}\right),
\end{equation}
which therefore provides a simple radially anisotropic DF for all the
doubloon models on the inner branch. For $p=-1$,
\begin{equation}
 F=\frac{
 \exp(\bar{\mathcal E})-3\exp(-\bar{\mathcal E})+2\exp(-2\bar{\mathcal E})}
 {(2\pi)^3GaL},
 \label{eq:jaffehalf}
\end{equation}
which is the DF for the Jaffe sphere.  This should be particularly
useful in setting up initial conditions for N-body experiments.

All the models on the inner branch exhibit an isothermal cusp
and so it is technically possible to set up the model composed
entirely of radial orbits, although
the resulting models will in general be prey to the radial
orbit instability \citep{Fr84,Pa87}.
The DFs of the radial orbit models are given in the closed form;
\begin{equation}
 F=\frac{V\delta(L^2)}{2^{3/2}\pi^3G}
 \biggl[
 \!\sqrt{|p|}\,(1+|p|)\,D_+\Bigl(\!\sqrt{|p|\bar{\mathcal E}}\Bigr)
 -\!\sqrt2|p|^\frac32D_+\Bigl(\!\sqrt{2|p|\bar{\mathcal E}}\Bigr)\biggr],
\label{eq:unstableDF}
\end{equation}
utilizing Dawson's integrals
\citep[see][though here we use $D_\pm$ instead of $F_\pm$ to avoid
confusion with the DF]{BT}:
\begin{equation}
 D_\pm(x) \equiv \exp(\mp x^2)\int_0^x\exp(\pm t^2)\,\dm t.
\end{equation}
This is actually closely related to the error function; that is,
$D_+(x)=(\!\sqrt\pi/2)\rme^{-x^2}{\rm erfi}(x)$ and
$D_-(x)=(\!\sqrt\pi/2)\rme^{x^2}{\rm erf}(x)$, where
${\rm erfi}(x)=\im^{-1}{\rm erf}(\im x)$ is the imaginary
error function. The radial velocity dispersions corresponding to
the pure radial orbit models are expressible using only elementary functions
\begin{equation}
\langle v_r^2\rangle={V^2\over |p|}
{(1+x)^2\log(1+x^{-1})-x-1-|p|/2\over 1+(1-|p|)\,x},
\end{equation}
where $x\equiv(a/r)^p=(r/a)^{|p|}$, which behaves like
$\langle v_r^2\rangle\simeq V^2\log(a/r)$ as $r\to0$ and
$\langle v_r^2\rangle\simeq V^2(2|p|)^{-1}(a/r)^{|p|}$
as $r\to\infty$, respectively.

\subsubsection{The Jaffe model}

In fact, the Jaffe model deserves particular attention,
as it is often used a model of a dark halo or an elliptical galaxy
\citep[see e.g.,][]{Ko96, Ge98}. The radial velocity dispersion
of the constant anisotropy Jaffe model is given by
\begin{subequations}\begin{equation}
 \langle v_r^2\rangle = V^2{ r^{2(1-\beta)}(a+r)^2 \over a^{2(2-\beta)}}
 {\rm B}_{a\over a+r}\bigl(5-2\beta,2\beta-2\bigr),
\end{equation}
which is reducible to an elementary function if $2\beta$ is an integer
-- in particular, if $n=2-2\beta$ is a non-negative integer,
\begin{equation}\begin{split}
 \frac{\langle v_r^2\rangle}{V^2} =
 { (1+n)(2+n) \over 2} \biggl(1+{r\over a}\biggr)^2
 \Phi_n\biggl({r\over a}\biggr) - { 3+n \over 2} -{2+n\over 2} {r \over a};
 \\ \Phi_n(x) = \sum_{k=1}^\infty{(-1)^{k-1}\over (n+k)x^k}
 =(-x)^n \biggl[\log\Bigl(1+{1\over x}\Bigr)
 +\sum_{k=1}^n{(-1)^k\over kx^k}\biggr].
\end{split}\end{equation}\end{subequations}

An analytic DF of the Jaffe model with $\beta=1/2$ is already provided
in equation (\ref{eq:jaffehalf}).  Similar analytic DFs of the Jaffe
model may also be obtained for all half-integer values of the
anisotropy parameter.  For instance, if $\beta=-1/2$,
\begin{equation}
 F=\frac{\bigl(\rme^{\bar{\mathcal E}}-1\bigr)^3
 \bigl(9+7\rme^{-\bar{\mathcal E}}+4\rme^{-2\bar{\mathcal E}}\bigr)\,L}
 {(2\pi)^3Ga^3V^2}.
\end{equation}

For an integer value of the anisotropy parameter,
the constant anisotropy DF of the Jaffe model is expressible as
a finite sum over Dawson's integrals. In particular, the DF
with only radial orbits is
\begin{equation}
F=\frac{V\delta(L^2)}{2\pi^3G}\left[
 \!\sqrt2\,D_+\Bigl(\!\sqrt{\bar{\mathcal E}}\Bigr)
 -D_+\Bigl(\!\sqrt{2\bar{\mathcal E}}\Bigr)\right],
\end{equation}
whilst the Jaffe models with $-\beta=n\in\{0,1,\dotsc\}$:
\begin{multline}
 F=\frac{a^{-2(1+n)}V^{-(1+2n)}L^{2n}}{2^{n+5/2}\pi^3n!G}
 \sum_{j=-2}^{2(n+1)}\frac{(-1)^j\!j^{n+2}}{|j|^{1/2}}{2n+4\choose j+2}\,
 D_{-({\rm sgn}\,j)}\Bigl(\!\sqrt{|j|\bar{\mathcal E}}\Bigr)
 \\
 =\frac{a^{-2(1+n)}V^{-(1+2n)}L^{2n}}{2^{n+5/2}\pi^{5/2}n!G}
 \Biggl[\sum_{j=1}^{2(n+1)}\frac{(-1)^j\!j^{n+\frac32}}2{2n+4\choose j+2}\,
 \rme^{j\bar{\mathcal E}}{\rm erf}\bigl(\!\sqrt{j\bar{\mathcal E}}\bigr)
 \\
 -(-1)^n(n+2)\rme^{-\bar{\mathcal E}}
 {\rm erfi}\bigl(\!\sqrt{\bar{\mathcal E}}\bigr)
 +(-2)^n\!\sqrt2\rme^{-2\bar{\mathcal E}}
 {\rm erfi}\bigl(\!\sqrt{2\bar{\mathcal E}}\bigr)\Biggr]
\end{multline}
where $D_0=0$ and ${a\choose n}$ is the binomial coefficient.
The isotropic Jaffe model is included as the special case ($n=0$) 
\begin{equation}
 F = { D_+\bigl(\!\sqrt{2\bar{\mathcal E}}\bigr)
   - \!\sqrt2 D_+\bigl(\!\sqrt{\bar{\mathcal E}}\bigr)
   + D_-\bigl(\!\sqrt{2\bar{\mathcal E}}\bigr)
   - \!\sqrt2 D_-\bigl(\!\sqrt{\bar{\mathcal E}}\bigr) \over 2\pi^3 Ga^2V}
\end{equation}
which was first given by \citet{Ja83} and is also repeated in
\citet{BT}.

The Jaffe models given here all have constant anisotropy. They may be
contrasted with the models found by \cite{Me85a}, which have isotropic
centers and strongly radially anisotropic ($\beta\to1$) outer parts.
These are derived using the inversion introduced by \citet{Os79} and
popularised by \cite{Me85b}. Although the transition from isotropy to
radial anisotropy is desirable, the Osipkov-Merritt models
unfortunately provide rather too extreme radial anisotropy in the
outer parts.

\begin{figure*}
  \centering
  \includegraphics[width=6in]{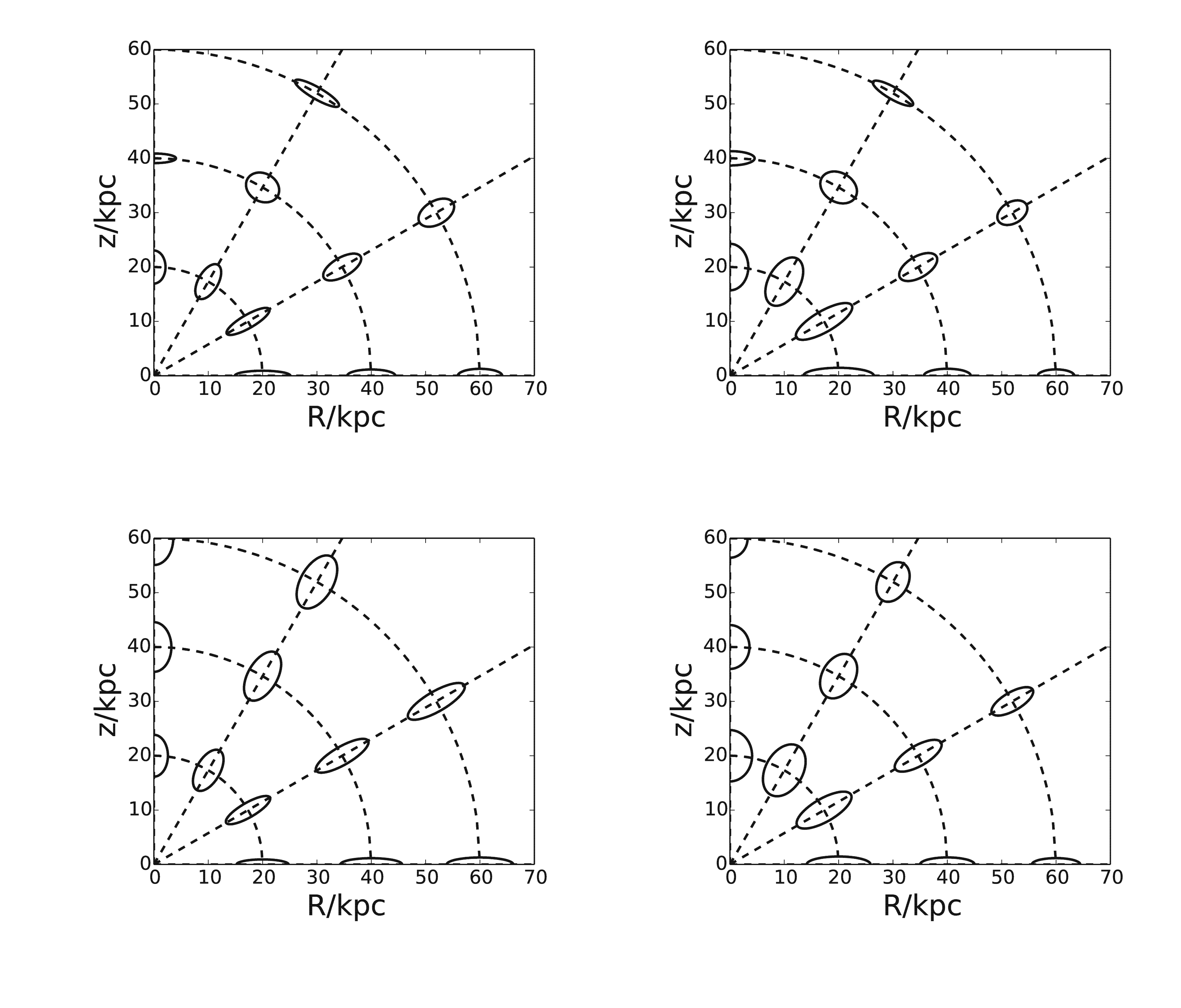}
  \caption{Velocity ellipsoids for flattened ($q=0.6$) Einasto
    profiles (upper panels) and power-law profiles (lower panels).
    The left column is for doubloon halos with $p = 1$ (outer branch),
    the right column for $p=-1$ (inner branch). All the doubloon
    models have $V = 200~\mbox{km s$^{-1}$}$ and $a = 50~\mbox{kpc}$.
    The Jeans solutions correspond to the case $K=1/40$. Notice that the
    flattened Einasto profiles do not have approximately invariant
    shapes, as the long axis of the velocity ellipsoid changes from
    radial to azimuthal at high latitudes. }
  \label{fig:velellipsoids}
\end{figure*}

\section{Flattened Tracer Populations: Jeans Equations}

For applications in which dark haloes are represented as doubloon
models, we are primarily interested in the properties of flattened
tracer populations of stars, whose kinematics are accessible to
observation. The flattening is usually described by assuming a density
law stratified on similar concentric spheroids with an axis ratio $q$.
Tracers in haloes are often modelled with power-laws or
broken power-laws \citep[e.g.,][]{Wa09,De11},
which have asymptotic behaviour
\begin{equation}
 \rho_\mathrm{pl} (r,\theta) \simeq Am^{-\delta},
 \quad m^2 = r^2(\sin^2\theta + q^{-2}\cos^2\theta).
\end{equation}
Another commonly used profile for luminous tracers,
whether in stellar haloes or elliptical galaxies,
is that due to J. Einasto \citep[see e.g.,][]{Ei89,De11},
though it has also been used for the dark matter distribution \citep{Gr06}.
It has asymptotic behaviour
\begin{equation}
\rho_\mathrm{Ein} (r,\theta) \simeq A\exp(-sm^{1/n}),
\end{equation}
where $s$ and $n$ are constants. This is the exponential profile
if $n=1$ and the de Vaucouleurs-like profile if $n=4$.
A convenient generalization that incorporates both families
is the hyper-Einasto profile \citep[cf.][]{AZ13}
\begin{equation}
\rho_\mathrm{HyEin} (r,\theta) \simeq Am^{-\delta}\exp(-sm^{1/n}).
\label{eq:HyEin}
\end{equation}
We shall solve the Jeans equations for both flattened power-law and
hyper-Einasto tracers in dark matter haloes described by doubloon
potentials.
In an axisymmetric model, $\langle v_r v_\phi \rangle \equiv \langle
v_\theta v_\phi \rangle \equiv 0$ by the symmetries of the individual orbits,
which then leaves two non-trivial Jeans equations
for the tracer density $\rho_{\rm t}(r,\theta)$
in terms of the spherical potential $\psi(r)$:
\begin{gather}\begin{split}
 {\upartial \rho_{\rm t} \langle v_r^2 \rangle \over \upartial r} 
 + {1 \over r} 
   {\upartial \rho_{\rm t} \langle v_r v_\theta \rangle \over \upartial \theta}
 + {\rho_{\rm t} \over r } \Bigl( 2\langle v_r^2 \rangle 
             - \langle v_\theta^2 \rangle 
             - \langle v_\phi^2 \rangle 
             + \langle v_r v_\theta \rangle \cot\theta \Bigr)
 \\
   = \rho_{\rm t} {\upartial\psi \over \upartial r} 
   = - \rho_{\rm t} {V^2 r^{p-1} \over a^p + r^p},
\end{split}\\
 {\upartial \rho_{\rm t} \langle v_r v_\theta \rangle \over \upartial r} 
 + {1 \over r} 
   {\upartial \rho_{\rm t} \langle v_\theta^2 \rangle \over \upartial \theta}
 + {\rho_{\rm t} \over r} \Bigl[ 3\langle v_r v_\theta \rangle 
             + (\langle v_\theta^2 \rangle 
                - \langle v_\phi^2 \rangle) \cot\theta \Bigr] 
  = 0.
\end{gather}
These two relations between the four stresses $\rho \langle v_r^2
\rangle$, $\rho \langle v_r v_\theta \rangle$, $\rho \langle
v_\theta^2 \rangle$, $\rho \langle v_\phi^2 \rangle$ must be satisfied
at any point in the model.

There is now increasing evidence from the stellar halo of our Galaxy
that the velocity ellipsoid is spherically aligned. This was first
noted by \citet{Sm09a}, who studied the kinematics of halo subdwarfs
in the Sloan Digital Sky Survey Stripe 82, for which there is
multi-epoch and multi-band photometry permitting the measurement of
accurate proper motions. Subsequently, \citet{Bo10} used $\sim 53000$
halo stars with $r$ band magnitude brighter than 20 and proper motion
measurements derived from Sloan Digital Sky Survey and Palomar
Observatory Sky Survey astrometry to extend this result over a quarter
of the sky at high latitudes.  Although the tilt of the velocity
ellipsoid in elliptical galaxies is not known, galaxy modelling
suggests that it is aligned on spheroidal coordinates
\citep[e.g.,][]{Bi14}, which become spherical at large radii.  In
other words, there is much motivation for investigating spherical
alignment, $\langle v_rv_\theta \rangle =0$, as it holds good for the
Milky Way's stellar halo and for the outer parts of elliptical
galaxies.

Additionally, there is good evidence from the kinematics of stars in
the Milky Way's stellar halo $\langle v_\theta^2 \rangle \approx
\langle v_\phi^2 \rangle$.  \citet{Sm09b} found $\langle v_\theta^2
\rangle^{1/2} = 82 \pm 2~\mbox{km s$^{-1}$}$ and $\langle v_\phi^2
\rangle^{1/2} = 77 \pm 2~\mbox{km s$^{-1}$}$.  \citet{Bo10} claimed
that the semi-axes are invariant over the volume probed by their much
larger sample and found $\langle v_\theta^2 \rangle^{1/2} = 85 \pm
5~\mbox{km s$^{-1}$}$ and $\langle v_\phi^2 \rangle^{1/2} = 75 \pm
5~\mbox{km s$^{-1}$}$.  It is interesting that the two angular
semi-axes are almost the same.  If the tracer population has a DF
depending on $E$ and $L$ only, then $\langle v_\theta^2 \rangle =
\langle v_\phi^2 \rangle$.  We accordingly make this assumption, which
has good theoretical and observational motivation, so that the Jeans
equations become
\begin{equation}
 {\upartial (r^2\rho_{\rm t} \langle v_r^2 \rangle) \over \upartial r} 
  = r \rho_{\rm t}
  \Bigl( 2\langle v_\theta^2 \rangle - v_{\rm c}^2 \Bigr), \quad
   {\upartial (\rho_{\rm t} \langle v_\theta^2 \rangle) \over \upartial \theta} 
   = 0.
\end{equation}
We see that this set of assumptions has closed the Jeans equations,
which may now be integrated with suitable boundary conditions.
Integrating the angular equation, we find $\rho_{\rm t} \langle
v_\theta^2 \rangle$ being an arbitrary function of $r$. As the
boundary condition, we next assume that the velocity dispersion on the
equatorial plane ($\theta = \pi/2$) is a constant fraction $K$ of the
rotation curve, that is,
\begin{equation}
\langle v_\theta^2 \rangle = \langle v_\phi^2 \rangle 
= K v_{\rm c}^2(r)\,  {\rho_{\rm t}(r, \pi/2) \over \rho_{\rm t}(r,\theta)}
\end{equation}
Assuming that $\rho\langle v_r^2 \rangle \to 0$ as $r\to\infty$,
we then have the full axisymmetric solution of the Jeans equations as
\begin{equation}
\langle v_r^2 \rangle = {1\over r^2 \rho_{\rm t}(r,\theta)}
\int_r^\infty\!\dm x\, x v_{\rm c}^2(x) \, 
\Bigl[\rho_{\rm t}(x,\theta) - 2K\rho_{\rm t}(x,\pi/2) \Bigr]
\end{equation}
This gives a one-parameter family of solutions of the Jeans equations
for axisymmetric densities with spherically aligned velocity
dispersion tensors. Algorithms for cylindrically aligned Jeans
solutions are known \citep{Ca08}, although as \citet{Bi14} points out
they are somewhat contrived. In cosmogonies in which galaxies are
built from hierarchical merging, stellar material on nearly radial
orbits fell in to deepening dark matter potential wells, and so
spherically aligned Jeans solutions are much more natural. A general,
though rather complicated, algorithm for spherically aligned Jeans
solutions for flattened potential--density pairs is known \citep{Ba83,Ba85}.
The assumption of a spherical potential, though, makes our
algorithm much simpler for flattened tracer densities.

A necessary condition for everywhere positive stresses is that
$0 < K < 1/2$. In practice, the range of physical $K$ values is much more
constrained, though established easily enough by numerical
integration. Fig.~\ref{fig:velellipsoids} shows velocity ellipsoids
for power-law and Einasto profiles representing the stellar halo of
the Milky Way. The Einasto profile has $n = 1.7$ and an effective
radius of $20~\mbox{kpc}$, the power law profile falls with $\delta = -4$
beyond a core radius of $0.6~\mbox{kpc}$. Both are inspired by fits to the
stellar halo of the Milky Way discussed in \citet{De11,De14} and
\citet{Ev14}. The left column shows each tracer in a doubloon model
with $p=-1$ \citep{Ja83}, the right with $p=1$ \citep{Ev14}.  It is
interesting to observe that the shape of the velocity ellipsoids is
primarily controlled by the tracer density, with the power-law or
Einasto profiles each generating similar Jeans solutions in different
doubloon potentials. However, the shape of the velocity ellipsoids for
power-laws tracers always has the radial velocity dispersion exceeding
the angular velocity dispersions, so that the velocity ellipsoids are
always prolate spheroids. This is not the case for the Einasto
profiles, in which the azimuthal velocity dispersions exceeds the
radial on approach to the poles ($\theta =0$), and so changes from
prolate to oblate spheroidal in shape.

Although \citet{De11,De14} found either power-law or Einasto profiles
equally good fits to the starcount data, it is obvious that the
kinematics provides a powerful discriminant. The fact that the
velocity ellipsoid shape is spherically aligned \citep{Sm09a} and
(to first order) shape invariant over the Sloan Digital Sky Survey
footprint \citep{Bo10} seems to rule out Einasto profiles. We plan to
return to detailed Jeans solution fits to the kinematics of the
stellar halo in a later publication.

\begin{figure*}
  \centering
  \includegraphics[width=3in]{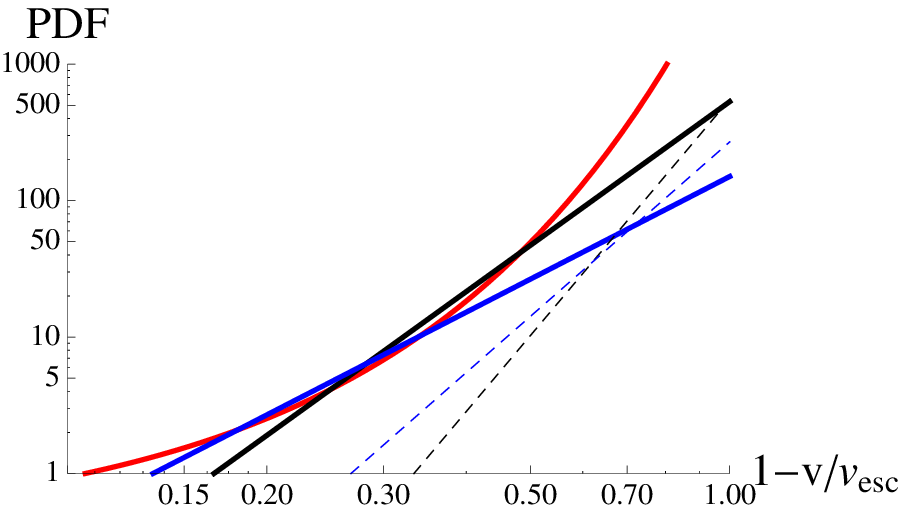}
  \includegraphics[width=3in]{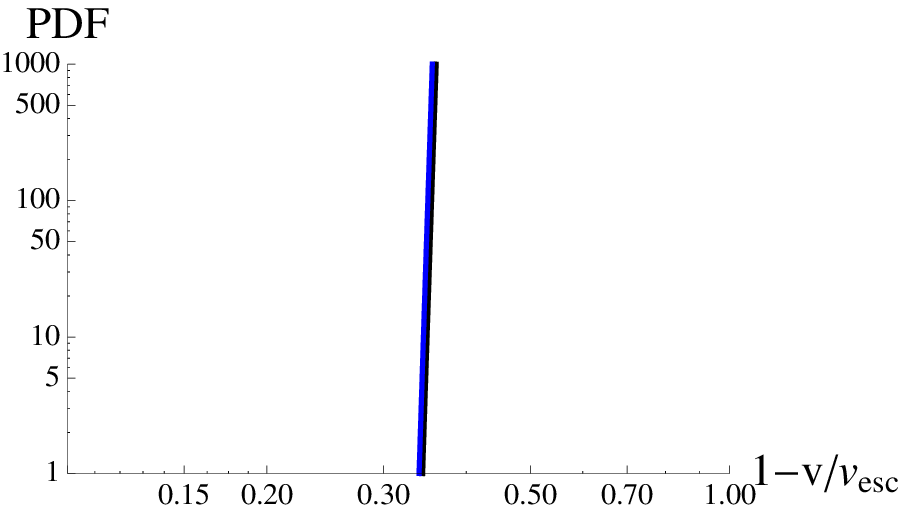}
  \caption{The probability density function of radial velocities near
    the escape speed for power-law (left) and Einasto (right) tracer
    densities. The logarithm of the probability density is plotted
    against $1 - v/v_{\rm esc}$. The red line is for an isotropic
    power-law tracer with $\delta = 5$ in a doubloon potential with
    $p=1$ (outer branch, the Evans \& Williams (2014) model). The
    black and blue lines show tracers in a doubloon potential with
    $p=-1$ (inner branch, Jaffe model) with $\delta =5$ and $4$. Full
    and dashed lines show isotropic ($\beta=0$) and radially
    anisotropic ($\beta=0.5$) models.  Although all four models are
    shown in the right panel, the super-exponential cut-off makes them
    virtually indistinguishable.}
  \label{fig:escapevel}
\end{figure*}

\section{Tracer Populations: Distributions of High Velocity Stars}

The high velocity stars of the Milky Way are distinct from the
hyper-velocity stars. The central black hole may eject
stars~\citep{Hi88, Yu03, Le06}, which are often unbound and moving on
highly radial orbits \citep{Br14}. These are the hyper-velocity stars,
which are a separate population and do not form part of the
steady-state stellar halo.

By contrast, the high velocity stars are the highest energy, but
bound, members of the halo.  The form of the distribution function at
the highest energies is accessible to observational scrutiny and can
in principle provide information on the behaviour of the potential at
the edge. For example, in the Milky Way, the distribution of
high-velocity stars from the halo is already available locally thanks
to the RAVE survey \citep{Sm07,Pi14,Ha14}. One of the earliest
investigations into the escape speed, \citet{Le90} introduced the
simple and attractive ansatz that
\begin{equation}\begin{cases}
f(v) \propto (v_{\rm esc} -v)^C & (v< v_{\rm esc}) \\
f(v) = 0                & (v> v_{\rm esc}),\end{cases}
\label{eq:lt}
\end{equation}
where $f(v)$ is the distribution of space velocities near the escape
speed $v_{\rm esc}$ and $C$ is a constant.  This ansatz, which is
exact for isotropic power-law DFs, has held up surprisingly well over
the last quarter of a century \citep{Pi14}. However, the next few
years will see the Gaia-ESO and LAMOST surveys, as well as the Gaia
satellite, substantially improve our knowledge of the distribution of
high velocity stars as a function of distance within 20~kpc of the
Sun. Sample sizes of hundreds or even thousands of high velocity stars
will become available, and deviations from equation (\ref{eq:lt}) can
be probed. Accordingly, we proceed to derive the form of the velocity
distribution at the highest energies for power-law and hyper-Einasto
tracers (defined in eq.~\ref{eq:HyEin}).

\subsection{Outer Branch}

The highest velocity stars in the outer branch models correspond
to the limit $E\to-\infty$, and we find
$r(E)\simeq a\rme^{|E|/V^2}\to\infty$ in the same limit.
If the tracer density asymptotically becomes a power-law like
$\rho\simeq Ar^{-\delta}$ as $r\to\infty$, then for $E\to-\infty$,
\begin{equation}
g\simeq 2^\beta Ar^{2\beta-\delta}
\simeq2^\beta Aa^{2\beta-\delta}
\exp\biggl(\frac{(\delta-2\beta)E}{V^2}\biggr).
\end{equation}
We find that the asymptotic form of the constant anisotropy DF is
\begin{equation}
F \simeq \frac{2^\beta A}{(2\pi)^{3/2}a^{\delta-2\beta}}
\biggl(\frac{\delta-2\beta}{V^2}\biggr)^{\frac32-\beta}
{L^{-2\beta}\over \Gamma(1-\beta)}
\exp\biggl(-\frac{(\delta-2\beta)|E|}{V^2}\biggr)
\label{eq:iso}
\end{equation}
for $E\to-\infty$.  So, an isotropic power-law tracer ($\beta=0$) has
an isothermal or a Maxwellian distribution of velocities. Even in the
presence of anisotropy, the distribution of radial velocities remains
a Maxwellian. The red line in the left panel of
Fig.~\ref{fig:escapevel} shows the probability density function (PDF)
of radial velocities near the escape speed derived from
eq.~(\ref{eq:iso}) for the model with $\delta =5$ and $\beta=0$.

If the tracer population has an hyper-Einasto profile,
then $g\simeq 2^\beta Ar^{2\beta-\delta}\exp(-sr^{1/n})$ as $r\to\infty$
and the DF asymptotically becomes
\begin{multline}\label{eq:pos}
F \simeq\frac{2^\beta A}{(2\pi)^{3/2}a^{\delta-2\beta}}
\biggl(\frac{sa^{1/n}}{nV^2}\biggr)^{\frac32-\beta}
{L^{-2\beta}\over \Gamma(1-\beta)}
\\\times\exp\Biggl[-sa^\frac1n\rme^{\frac{|E|}{nV^2}}
-\biggl(\delta-2\beta-\frac{3-2\beta}{2n}\biggr)\,\frac{|E|}{V^2}\Biggr]
\end{multline}
as $E\to-\infty$.  The distribution of space velocities is no longer
Maxwellian, but rather is a Maxwellian modulated by a
super-exponential cut-off.

\subsection{Inner Branch}

The forms of the high energy tail of the velocity distribution change
if the dark matter density falls off like $r^{-3}$ or faster, as we
now show. For $p<0$, Taylor expansion now shows that $r(E)\simeq
a(V^2/|p|)^{1/|p|}E^{-1/|p|}\to\infty$ in the limit $E\to0$.  For
power-law tracers, then as $r\to\infty$ and $E\to0$,
\begin{equation}
g\simeq 2^\beta Ar^{2\beta-\delta}
\simeq2^\beta Aa^{2\beta-\delta}
\biggl(\frac{|p|E}{V^2}\biggr)^\chi
\quad(\text{where }\ \chi\equiv\frac{\delta-2\beta}{|p|})
\end{equation}
which is also a power-law and so the asymptotic form of the DF is
\begin{equation}
F \simeq\frac{2^\beta A}{(2\pi)^{3/2}a^{\delta-2\beta}}
\frac{\Gamma(\chi+1)}
{\Gamma(\chi-1/2+\beta)}
{L^{-2\beta}\over \Gamma(1-\beta)} E^{\beta-\frac32}
\biggl(\frac{|p|E}{V^2}\biggr)^\chi
\label{eq:pow}
\end{equation}
for $E\to0$. In other words, the space velocity distribution of a
power-law tracer falls asymptotically like a power-law at the highest
energies. The black and blue lines in the left panel of
Fig.~\ref{fig:escapevel} show the PDF of radial velocities derived
from equation (\ref{eq:pow}) for the model with $\delta =5$ and
$\delta =4$.  Full lines are isotropic ($\beta=0$), dotted lines
radially anisotropic ($\beta = 1/2$).

For an Einasto tracer, a similar calculation yields
\begin{multline}\label{eq:neg}
F\simeq
\frac{2^\beta A}{(2\pi)^{3/2}a^{\delta-2\beta}}
{L^{-2\beta}\over \Gamma(1-\beta)}
\\\times
\biggl(\frac{sa^{1/n}}{n|p|E}\biggr)^{\frac32-\beta}
\biggl(\frac{|p|E}{V^2}\biggr)^{\frac{\delta-2\beta}{|p|}-\frac{3-2\beta}{2n|p|}}
\exp\Biggl[-sa^\frac1n\biggl(\frac{V^2}{|p|E}\biggr)^\frac1{n|p|}\Biggr]
\end{multline}
as $E\to0$. So, the distribution of space velocities falls like a
power-law with a super-exponential cut-off. The same four models
($\delta =5$ and $4$, and $\beta =0$ and $1/2$) are shown in the right
panel of Fig.~\ref{fig:escapevel}. The super-exponential cut-off
causes all four model to lie on top of each other.

\section{Conclusions}

We have presented details of a new family of spherical models, which
have properties suitable for mimicking galaxies with flattish rotation
curves. The models may have a rotation curve which attains a finite
value at the centre and falls on moving outwards. The archetype is the
\citet{Ja83} model. Alternatively, the models can have a rising
rotation curve in the inner parts, which flattens
asymptotically. Here, the archetype is the spherical logarithmic model
popularised by \citet{BT}. The family also includes the singular
isothermal sphere with an $\rho \sim r^{-2}$ density cusp at the
center, as well as the halo model recently discovered by \citet{Ev14}
which has an $r^{-1}$ cusp.  In general, the family includes both
cored and cusped members, and so can represent the range of cusps
found in numerical simulations~\citep[see e.g.,][]{Mo99}.

The halo models presented here are spherical. Cosmological simulations
suggest that dark halos are typically flattened and mildly oblate,
with a ratio of long axis to short axis of $q=0.8$-$0.9$~\citep[see
  e.g.,][]{Ab10, De11, Ze12}. For the Milky Way Galaxy, there are
several lines of evidence suggesting that the dark halo may be nearly
spherical. First, the fits to the tidal stream GD-1 using different
methods~\citep{Ko10,Bo14} show that the total Galactic potential (disk
and halo) at radii $\sim 15$ kpc is consistent with modest flattening.
Secondly, the kinematics of halo subdwarfs in the Sloan Digital Sky
Survey Stripe 82, for which there is multi-epoch and multi-band
photometry permitting the measurement of accurate proper motions, is
consistent with a nearly spherical potential~\citep{Sm09a}.  Thirdly,
the kinematics of the Sagittarius (Sgr) stream assuredly prohibit
strongly flattened dark haloes with $q<0.6$~\citep{EvB14}. Whilst a
definitive answer from Sgr stream kinematics must await a thorough
exploration of stream generation in flattened haloes using modern
algorithms~\citep{Gi14}, the debris of the Sgr stretches a full
$360\degr$ over the sky and is almost confined to a plane.

The self-consistent distribution functions (DFs) in terms of the
energy $E$ and angular momentum $L$ have been given for the models
with constant anisotropy. There are however strong reasons for using
actions, instead of integrals of the motion like energy, in
DFs~\citep{Bi13}. Here, we note that \citet{WE14} have shown that the
Hamiltonian as a function of actions in scale-free power-laws is an
almost linear function of the actions, enabling schemes to be easily
devised to convert the DFs into action space if desired. \citet{Ev14}
provide a practical example of such an algorithm for one member of the
doubloon family. \citet{Po14} and \citet{Wi14} give algorithms for
action-based DFs for models with density falling like power-laws in
certain regimes.

The self-consistent DFs describe the velocity distributions of dark
matter needed to sustain the doubloon models. For applications to
stars in stellar halos or in elliptical galaxies, we are interested in
the properties of luminous tracers within the doubloon models. We have
provided a number of results for both power-law and Einasto tracer
populations. There is good evidence that the velocity dispersion
tensor of the Milky Way's stellar halo is aligned in spherical polar
coordinates~\citep{Sm09a, Bo10}. Additionally, the velocity ellipsoid
of stellar populations in elliptical galaxies is probably aligned in
spheroidal coordinates, which asymptotically become
spherical. Therefore, spherically aligned Jeans solutions are of
considerable astrophysical interest. We identify a simple algorithm
for solving the spherically aligned Jeans equations for flattened
tracers in spherical potentials, and provide solutions for power-law
and hyper-Einasto tracers (defined in eq.~\ref{eq:HyEin}).

The distribution of high velocity stars of the stellar halo of the
Milky Way are becoming available~\citep{Sm07, Pi14, Ha14}.  So, we
have provided the asymptotic forms of the DFs for tracer populations
with power-law and Einasto density distributions.  The form of the
high velocity tail is particularly interesting as this may betray
properties of the dark matter halo. Power-law tracers have velocity
distributions with power-law or Maxwellian tails. Einasto tracers have
super-exponential cut-offs in the velocity distributions. Although the
observational data are often fitted to models like $f(v) \propto
(v_{\rm esc} -v)^k$ this is only strictly correct for power-law tracers. If
the stellar halo is described by an Einasto profile, then

\begin{equation}
\begin{cases} f(v) \propto (v_{\rm esc} -v)^C \exp \Bigl[ -{\displaystyle A\over \displaystyle (v_{\rm esc}-v)^B} \Bigl] & v< v_{\rm esc} \\
               f(v)    =0 & v> v_{\rm esc}, \end{cases}
\end{equation}
with $A, B$ and $C$ constants, is a better description of the high
velocity tail near the escape speed.  In principle, the different
forms of the velocity distributions of tracers can provide us with
evidence on the extent of the dark matter potential.  This is a
subject on which there is not merely little hard evidence, but very
few avenues in which to gain evidence.

\section*{Acknowledgments}
We thank the anonymous referee for a very useful report.
AB and AW are supported by the Science and Technology Facilities
Council (STFC) of the United Kingdom. JA is supported by the Chinese
Academy of Sciences (CAS) Fellowships for Young International Scientists
(Grant No.~2009YAJ7) and also
grants from the National Science Foundation of China (NSFC).

\bibliographystyle{mn2e}

\begin{thebibliography}{99}

\bibitem[Abadi et al. (2010)]{Ab10} Abadi M. G., Navarro J. F., Fardal
    M., Babul A., Steinmetz M., 2010, MNRAS, 407, 435

\bibitem[Abramowitz \& Stegun(1964)]{AS} Abramowitz M., Stegun I., 1964,
  Handbook of Mathematical Functions,
  National Bureau of Standards, Washington DC
  ({\sc reprinted} 1972, Dover, New York)

\bibitem[An \& Evans(2006a)]{An06} An J. H., Evans N. W., 2006a,
  ApJ, 642, 752 

\bibitem[An \& Evans(2006b)]{An06b} An J. H., Evans N. W., 2006b,
  A\&A, 444, 45

\bibitem[An \& Zhao(2013)]{AZ13} An J., Zhao H., 2013, MNRAS, 428, 2805

\bibitem[Bacon et al.(1983)]{Ba83} Bacon R., Simien F., Monnet G., 1983,
  A\&A, 128, 405

\bibitem[Bacon(1985)]{Ba85} Bacon R., 1985, A\&A, 143, 84 

\bibitem[Binney(1981)]{Bi81} Binney J. J., 1981, MNRAS, 196, 455

\bibitem[Binney(2013)]{Bi13} Binney J. J., 2013, New Astron.~Rev., 57, 29

\bibitem[Binney(2014)]{Bi14} Binney J., 2014, MNRAS, 440, 787 

\bibitem[Binney \& Tremaine(2008)]{BT} Binney J., Tremaine S., 2008,
  Galactic Dynamics, 2nd edn. Princeton University Press, Princeton

\bibitem[Bond et al.(2010)]{Bo10} Bond N. A., Ivezi{\'c} {\v Z}.,
  Sesar B., et al., 2010, ApJ, 716, 1

\bibitem[Bowden et al.(2014)]{Bo14} Bowden A., Belokurov V., Evans
  N. W., 2014, MNRAS, arXiv1502.00484

\bibitem[Brown et al.(2014)]{Br14} Brown W. R., Geller M. J., Kenyon S. J., 2014, ApJ, 787, 89

\bibitem[Cappellari(2008)]{Ca08} Cappellari M., 2008, MNRAS, 390, 71

\bibitem[Chandrasekhar(1939)]{Ch10} Chandrasekhar S., 1939,
  An Introduction to the Study of Stellar Structure,
  the University of Chicago Press, Chicago
  ({\sc reprinted} 1958, {\sc reissued} 2010, Dover, New York)

\bibitem[Deason et al.(2011)]{De11} Deason A. J., Belokurov V.,
  Evans N. W., 2011, MNRAS, 416, 2903

\bibitem[Deason et al.(2011)]{De11a} Deason A. J.,
  et al., 2011, MNRAS, 415, 2607

\bibitem[Deason et al.(2014)]{De14} Deason A. J., Belokurov V.,
  Koposov S. E., Rockosi C. M., 2014, ApJ, 787, 30

\bibitem[Dehnen(1993)]{De93} Dehnen W., 1993, MNRAS, 265, 250

\bibitem[Eddington(1916)]{Ed16} Eddington A. S., 1916, MNRAS, 76, 572

\bibitem[Einasto \& Haud (1989)]{Ei89} Einasto J., Haud U., 1989, A\&A, 223, 89

\bibitem[Erd\'elyi et al.(1953)]{Er53} Erd\'elyi A., Magnus W.,
  Oberhettinger F., Tricomi F., 1953, Higher Transcendental Functions,
  McGraw-Hill, New York
  
\bibitem[Erd\'elyi et al.(1954)]{Er54} Erd\'elyi A., Magnus W.,
  Oberhettinger F., Tricomi F., 1954, Tables of Integral Transforms,
  McGraw-Hill, New York

\bibitem[Evans(1993)]{Ev93} Evans N. W., 1993, MNRAS, 260, 191

\bibitem[Evans \& An(2005)]{Ev05} Evans N. W., An J., 2005, MNRAS, 360, 492

\bibitem[Evans \& An(2006)]{Ev06} Evans N. W., An J. H., 2006, Phys. Rev.~D, 73, 023524 

\bibitem[Evans \& Bowden(2014)]{EvB14} Evans N. W., Bowden A., 2014, MNRAS, 443, 2

\bibitem[Evans \& Williams(2014)]{Ev14} Evans N. W., Williams A. A., 2014, MNRAS, 443, 791 

\bibitem[Fridman \& Polyachenko(1984)]{Fr84} Fridman A. M., Polyachenko
  V., 1984, Physics of Gravitating Systems, Springer Verlag, Berlin

\bibitem[Gerhard et al.(1998)]{Ge98} Gerhard O. E., Jeske G., Saglia
  R. P., Bender R. 1998, MNRAS, 295, 197

\bibitem[Gibbons et al(2014)]{Gi14} Gibbons S., Belokurov V., Evans N. W., 2014, MNRAS, 445, 3788

\bibitem[Graham et al.(2006)]{Gr06} Graham A. W., Merritt D., Moore B.,
  Diemand J., Terzic B., 2006, AJ, 132, 2701
 
\bibitem[Hansen \& Moore(2006)]{Ha06} Hansen S., Moore B., 2006, New Astron., 11, 333

\bibitem[Hawkins et al.(2015)]{Ha14} Hawkins K.,
et al., 2015, MNRAS, 447, 2046

\bibitem[Hernquist(1990)]{He90} Hernquist L., 1990, ApJ, 356, 359

\bibitem[Hernquist \& Ostriker(1992)]{He92} Hernquist L., Ostriker J. P., 1992, ApJ, 386, 375 

\bibitem[Hills(1988)]{Hi88} Hills J. G., 1988, Nature, 331, 687

\bibitem[Iguchi et al.(2006)]{Ig06} Iguchi O., Sota Y., Nakamichi A.,
  Morikawa M. 2006, Phys Rev E, 73, 046112

\bibitem[Jaffe(1983)]{Ja83} Jaffe W., 1983, MNRAS, 202, 995

\bibitem[Jeans(1919)]{Je19} Jeans J. H., 1919, Problems of Cosmogony and
  Stellar Dynamics, Cambridge University Press, Cambridge

\bibitem[Kochanek(1996)]{Ko96} Kochanek C., 1996, ApJ, 457, 228

\bibitem[Koposov et al.(2010)]{Ko10} Koposov S., Rix H. W., Hogg D., 2010, ApJ, 712, 260

\bibitem[Leonard \& Tremaine(1990)]{Le90} Leonard P. J. T., Tremaine
  S., 1990, ApJ, 353, 486

\bibitem[Levin(2006)]{Le06} Levin Y., 2006, ApJ, 653, 1203 

\bibitem[Merritt(1985a)]{Me85a} Merritt D., 1985a, MNRAS, 
214, 25P 

\bibitem[Merritt(1985b)]{Me85b} Merritt D., 1985b, AJ, 90, 
1027 

\bibitem[Mo, van den Bosch \& White(2010)]{Mo10} Mo H., van den Bosch
  F., White S. D. M., 2010, Galaxy Formation and Evolution, Cambridge
  University Press, Cambridge

\bibitem[Moore et al.(1999)]{Mo99} Moore B., Quinn T., Governato F., Stadel J., Lake G., 1999, MNRAS, 310, 1147 

\bibitem[Olver et al.(2010)]{DLMF}
Olver F. W. J., Lozier D. W., Boisvert R. F., Clark C. W., 2010,
NIST Handbook of Mathematical Functions. Cambridge University Press, Cambridge
(\url{http://dlmf.nist.gov})

\bibitem[Osipkov(1979)]{Os79} Osipkov, L. P., 1979, Pisma Astron. Zh., 5, 77 

\bibitem[Palmer \& Papaloizou(1987)]{Pa87} Palmer P. L, Papaloizou J. C. B., 1987, MNRAS, 224, 1043

\bibitem[Piffl et al.(2014)]{Pi14} Piffl T.,
et al., 2014, A\&A, 562, A91

\bibitem[Plummer(1911)]{Pl11} Plummer H. C., 1911, MNRAS, 71, 460

\bibitem[Posti et al.(2015)]{Po14} Posti L., Binney J., Nipoti C.,
  Ciotti L., 2015, MNRAS, 447, 3060

\bibitem[Sota et al.(2008)]{So08} Sota Y., Iguchi O., Tashiro T.,
  Morikawa M.. 2008, Phys Rev E, 77, 05117

\bibitem[Smith et al.(2007)]{Sm07} Smith M. C.,
et al., 2007, MNRAS, 379, 755 

\bibitem[Smith et al.(2009a)]{Sm09a} Smith M. C., Evans N. W., An J. H., 2009a, ApJ, 698, 1110

\bibitem[Smith et al.(2009b)]{Sm09b} Smith M. C.,
et al., 2009b, MNRAS, 399, 1223 

\bibitem[Tremaine et al.(1994)]{Tr94} Tremaine S., Richstone D. O., Byun Y.-I., et al., 1994, AJ, 107, 634 

\bibitem[Watkins et al.(2009)]{Wa09} Watkins L. L.,
et al., 2009, MNRAS, 398, 1757

\bibitem[Wilkinson \& Evans (1999)]{We99} Wilkinson M., Evans N. W., 1999, MNRAS, 310, 645
  
\bibitem[Williams \& Evans(2015)]{Wi14} Williams A. A., Evans N. W.,
  2015, MNRAS, 448, 1360

\bibitem[Williams et al.(2014)]{WE14} Williams A. A., Evans N. W., Bowden A., 2014, MNRAS 442, 1405


\bibitem[Yu \& Tremaine(2003)]{Yu03} Yu Q., Tremaine S., 2003, ApJ, 599, 1129 

\bibitem[Zemp et al.(2012)]{Ze12} Zemp M., Gnedin O. E., Gnedin N. Y., Kravtsov A. V., 2012, ApJ, 748, 54

\end{thebibliography}

\appendix

\section{Constant anisotropy models}
\label{app}

\subsection{Fractional calculus}

Many of the formulae concerning the constant anisotropy DF
can be swiftly derived using the notion of fractional differentiation.
Let us consider the Riemann--Lioville integral \citep[see][]{Er54}:
\begin{equation}
{^+_aI}_x^\mu f(x) \equiv \frac1{\Gamma(\mu)}
\int_a^x\!\dm y\,(x-y)^{\mu-1}f(y)\quad(\mu>0).
\end{equation}
This generalizes Cauchy's formula for repeated integration for 
an arbitrary positive order.
This is also recognized as (generalized) Abel transform for $0<\mu<1$
with the classical case resulting from $\mu=1/2$.
These obey the composition rule
\begin{equation}
{^+_aI}_x^\mu\,{^+_aI}_x^\xi f(x)={^+_aI}_x^{\mu+\xi}f(x),
\end{equation}
and the differentiation follows
\begin{equation}
\biggl(\frac\dm{\dm x}\biggr)^n{^+_aI}_x^\mu f(x)=
\begin{cases}{^+_aI}_x^{\mu-n}f(x)&(n<\mu)\\f(x)&(n=\mu)\end{cases}.
\end{equation}
Thus the Riemann--Lioville integral may be inverted through
\begin{equation}
g(x)={^+_aI}_x^\mu f(x)\Rightarrow\
f(x)=\biggl(\frac\dm{\dm x}\biggr)^{\lfloor\mu\rfloor+1}
{^+_aI}_x^{1-\epsilon}g(x)
\end{equation}
where $\lfloor\mu\rfloor$ and $\epsilon\equiv\mu-\lfloor\mu\rfloor$
are the integer floor and the fraction part of $\mu$, respectively.
Let us also define the fractional derivative:
\begin{equation}
{_a\upartial_x}^\xi f(x)
\equiv \frac{\Gamma(\xi+1)}{2\pi\im}
\int\limits_a^{(x+)}\frac{f(z)\,\dm z}{(z-x)^{1+\xi}}
\end{equation}
by means of the complex contour integral.
Here, the contour starts and ends at the base point $a$, and encircles
$x$ in the counter-clockwise direction.
Thanks to Cauchy's integral (and differentiation) formula,
this coincides with the customary notion of differentiation
for integer values of $\xi$, that is,
${_a\upartial_x}^0f(x)=f(x)$ and ${_a\upartial_x}^nf(x)=f^{(n)}(x)$
for $n\in\{1,2,\dotsc\}$. Explicit calculations can demonstrate that
the fractional derivative is the inverse operator of
the Riemann--Lioville integral, whereby
\begin{equation}
{_a\upartial_x}^\xi\,{^+_aI}_x^\xi f(x)=f(x),
\end{equation}
and so follows that
${_a\upartial_x}^\xi f(x)
=(\dm/\dm x)^{\lceil\xi\rceil}{^+_aI}_x^{\lceil\xi\rceil-\xi}f(x)$
where $\lceil\xi\rceil$ is the integer ceiling of $\xi$
and also assuming ${^+_aI}_x^0f(x)=f(x)$.

\subsection{Constant anisotropy DF}

Suppose that the DF is given by the ansatz of equation (\ref{eq:cdf}).
The density profile results from the integral over the velocity,
\begin{equation}\begin{split}
\rho&=\iiint\!\dm^3\!\bmath v\,F
=\frac{2\pi}{r^2}\!\iint_{E\ge a,L^2\ge 0,v_r^2\ge 0}\!
\frac{F(E,L^2)\,\dm E\,\dm L^2}{\sqrt{v_r^2(E,L^2;\psi,r)}}
\\&=(2r^2)^{-\beta}\,{^+_aI}_\psi^{\frac32-\beta}f_E(\psi),
\end{split}\end{equation}
where and $a\equiv\psi(r_0)$ if $r_0$ is the finite boundary radius or
$a\equiv\lim_{r\to\infty}\psi(r)$. Hence if we define the augmented density
\begin{equation}
g(\psi)\equiv(2r^2)^\beta\rho={^+_aI}_\psi^{\frac32-\beta}f_E(\psi),
\end{equation}
the energy part of the DF is inverted as
\begin{equation}\label{eq:feinv}
f_E(E)={_a\upartial_E}^{\frac32-\beta}g(E)
=\biggl(\frac\dm{\dm E}\biggr)^{n+1}{^+_aI_E}^{1-\epsilon}g(E),
\end{equation}
where $n=\lfloor(3/2-\beta)\rfloor$ and $\epsilon=3/2-\beta-n$, which
may be compared with equation (3) of \citet{Ev06} (note the scale
constants are chosen differently here). For the isotropic case
($\beta=0$), we have
\begin{equation}
(2\pi)^{3/2}F(E)={_a\upartial_\psi}^{\frac32}\rho(\psi)\bigr|_{\psi=E}
=\frac{\dm^2}{\dm E^2}{^+_aI_\psi}^\frac12\rho(\psi)\Bigr|_{\psi=E}
\end{equation}
which reproduces Eddington's (1916) formula, which can be thought of
as the fractional derivative of order $3/2$.

For the DF of the form of equation (\ref{eq:cdf}), the energy part
$f_E(E)$ is directly related to the local energy distribution
\begin{equation}
\frac{\dm\rho}{\dm E}
=\frac{[\psi(r)-E]^{1/2-\beta}}{\Gamma(3/2-\beta)}\frac{f_E(E)}{(2r^2)^\beta}
\end{equation}
as well as the speed distribution
\begin{equation}
\frac{\dm\rho}{\dm v^2}
=\frac{v^{1-2\beta}f_E[\psi(r)-v^2/2]}{2^{3/2}\Gamma(3/2-\beta)\,r^{2\beta}}.
\end{equation}
Here also note that the local escape speed is given by
$v_{\rm esc}=\sqrt{2\psi(r)}$ provided that $\lim_{r\to\infty}\psi(r)=0$.

\subsection{Auxiliary results}

For models on the outer branch, we make use of Hankel's Loop
integral for the reciprocal gamma function
(see \citealt[eq.1.6(2)]{Er53}; \citealt[\S~5.9(1)]{DLMF})
%
%
indicating
\begin{equation}\label{eq:hankel}
{_{-\infty}\upartial_E}^\lambda\rme^{sE}
=\frac{\Gamma(\lambda+1)}{2\pi\im}
\int\limits_{-\infty}^{(E+)}\frac{\exp(s\psi)\,\dm\psi}{(\psi-E)^{1+\lambda}}
=s^\lambda\rme^{sE}.
\end{equation}

For models on the inner branch however, we need to introduce the
incomplete gamma function.  The contour integral representation of the
lower incomplete gamma function (cf.\ \citealt[eq.9.3(1)]{Er53};
\citealt[\S~8.6(2)]{DLMF})
%
%
implies
\begin{equation}\label{eq:igc}\begin{split}
{_0\upartial_E}^\lambda\rme^{sE}
&=\frac{\Gamma(\lambda+1)}{2\pi\im}
\int\limits_0^{(E+)}\frac{\exp(s\psi)\,\dm\psi}{(\psi-E)^{1+\lambda}}
\\&=s^\lambda\rme^{sE}P(-\lambda,sE)
=E^{-\lambda}\rme^{sE}\gamma^*(-\lambda,sE)
\end{split}\end{equation}
Here, $P(a,x)\equiv\gamma(a,x)/\Gamma(a)=x^a\gamma^*(a,x)$ is the
regularized lower incomplete gamma function.
If $\lambda$ is a positive integer, then $P(-\lambda,x)=1$ and
so this is same as the ordinary $n$-th derivative.  For a half-integer
$\lambda$, this reduces to the error function
(or Dawson's integral, which is equivalent to the imaginary error integral)
-- note
${\rm erf}(x)=(1/\sqrt{\pi})\gamma(1/2,x^2)=P(1/2,x^2)$ and ${\rm erfi}(x)=-\im\,{\rm erf}(\im x)$.

\subsection{Constant anisotropy DFs of self-consistent Doubloons}

For models on the outer branch, $p>0$ and $\phi<0$ and so
$0<\varepsilon<1$ where $\varepsilon=\exp(p\psi/V^2)$.
Then $g(\psi)$ of equation (\ref{eq:doubloonad}) may be expanded
to the power-series of $\varepsilon$ using binomial series;
\begin{equation}\begin{split}
  g(\psi)
    &=\frac{C(\varepsilon+p\varepsilon^2)}{\varepsilon^{-\lambda}}
    \sum_{j=0}^\infty\frac{(\lambda)_j}{j!}\varepsilon^j
    \\&=C\left[\varepsilon^{\lambda+1}+
    \sum_{j=1}^\infty\biggl(\frac{(\lambda)_j}{j!}
    +\frac{p(\lambda)_{j-1}}{(j-1)!}\biggr)\,\varepsilon^{\lambda+1+j}\right],
\end{split}\end{equation}
where $\lambda\equiv2p^{-1}(1-\beta)-1$ and
$C=2^{\beta-2}a^{2(\beta-1)}V^2/(\pi G)$.
Equation (\ref{eq:hankel}) then indicates
${_{-\infty}\upartial_\psi}^{\frac32-\beta}\varepsilon^{\lambda+1+j}
=V^{2\beta-3}[p(\lambda+1+j)]^{\frac32-\beta}\varepsilon^{\lambda+1+j}$,
and so equation (\ref{eq:feinv}) results in the DF in the form
of equation (\ref{eq:cdf}) given by equation (\ref{eq:cdfo}).
For models on the inner branch, we have
\begin{equation}
g(\psi) = \tilde g(\rme^{-|p|\psi/V^2}),\quad
\tilde g(\varepsilon)
={C(1-|p|\varepsilon)(1-\varepsilon)^{\xi+1}\over \varepsilon^\xi},
\end{equation}
where $\xi=2|p|^{-1}(1-\beta)\ge0$. If $\beta=\frac32-n$,
equation (\ref{eq:feinv}) becomes
\begin{equation}
f_E(E)=\frac{\dm^ng(E)}{\dm E^n}
=\biggl(-\frac{|p|\varepsilon}{V^2}\frac\dm{\dm\varepsilon}\biggr)^n
\tilde g(\varepsilon)\Biggr\rvert_{\varepsilon=\exp(-|p|E/V^2)},
\end{equation}
which can be computed analytically. For others, we can still expand
$\tilde g(\varepsilon)$ in a power series in $\varepsilon$ (note
$0<\varepsilon<1$ since $p<0$ but $\psi>0$);
\begin{equation}
g(\psi)={C(1-|p|\varepsilon)\over \varepsilon^\xi}
\sum_{j=0}^\infty{\xi+1\choose j}(-\varepsilon)^j.
\end{equation}
Note, if $\xi$ is a non-negative integer, the sum terminates after
the $j=\xi+1$ term and so reduces to a polynomial in $\varepsilon$.
Given equation (\ref{eq:igc}), applying equation (\ref{eq:feinv})
then results in $f_E(E)$ given as a sum over incomplete gamma functions,
$P[\beta-\frac32,-(j-\xi)|p|\bar{\mathcal E}]$
where $\bar{\mathcal E}=E/V^2(\ge0)$. If $\beta$ is an integer,
these are reducible to the error function or Dawson's integral,
particular examples of which are provided in the main body.

\end{document}